\documentclass[11pt,twoside]{article}

\usepackage{asp2004}
\usepackage{epsf}
\usepackage{lscape}
\usepackage{natbib}

\begin{document}
\title{The Wolf-Rayet hydrogen puzzle -- an observational point of view}   
\author{C. Foellmi$^1$, S.V. Marchenko$^2$ \& A.F.J. Moffat$^3$}   
\affil{1. European Southern Observatory, 3107 Alonso de Cordova, 19 Vitacura, Santiago, Chile. ({\tt cfoellmi@eso.org}) \\ 2. Department of Physics and Astronomy, Thompson Complex Central Wing, Western Kentucky University, Bowling Green, KY 42101-3576, USA. \\ 3.D\'epartement de physique, Universit\'e de Montr\'eal, C.P. 6128, Succ. Centre-Ville, Montr\'eal, QC, H3C 3J7, Canada}   

\begin{abstract} 
Significant amounts of hydrogen were found in very hot early-type {\it single} WN stars in the SMC and the LMC. Recently, we found similar evidence in the Wolf-Rayet star of a short-period LMC {\it binary}. We discuss here the relevance of hydrogen for WR star classification, models, the relation to metallicity, and the GRB progenitors.
\end{abstract}

\section{The first hints}

Hydrogen is obviously a crucial element in stellar evolution, even in core helium-burning Wolf-Rayet (WR) stars. It has been found in the SMC that about half of the WR population consists of hot, single, hydrogen-containing {\it single} WN3 and WN4 stars \citep[classified WN3-4ha or (h)a,][]{Foellmi-etal-2003a,Foellmi-2004}. For such early spectral types, significant amounts of hydrogen were not expected, since such hot WN stars are believed to have peeled off their outer H-rich layers by a very strong and optically thick wind, combined with internal convective mixing to expose core-processed material. Contrary to expectations, blue-shifted absorption lines of HI and HeII were detected in the spectra of all single SMC WNE stars, and Foellmi et al. argued that they originate in a WR wind. Given their dominance in the SMC, it is likely that these stars have a low initial mass (say 25-40 $M_{\odot}$) and are formed mainly thanks to high rotational velocity.

\begin{figure}[!ht]
\plottwo{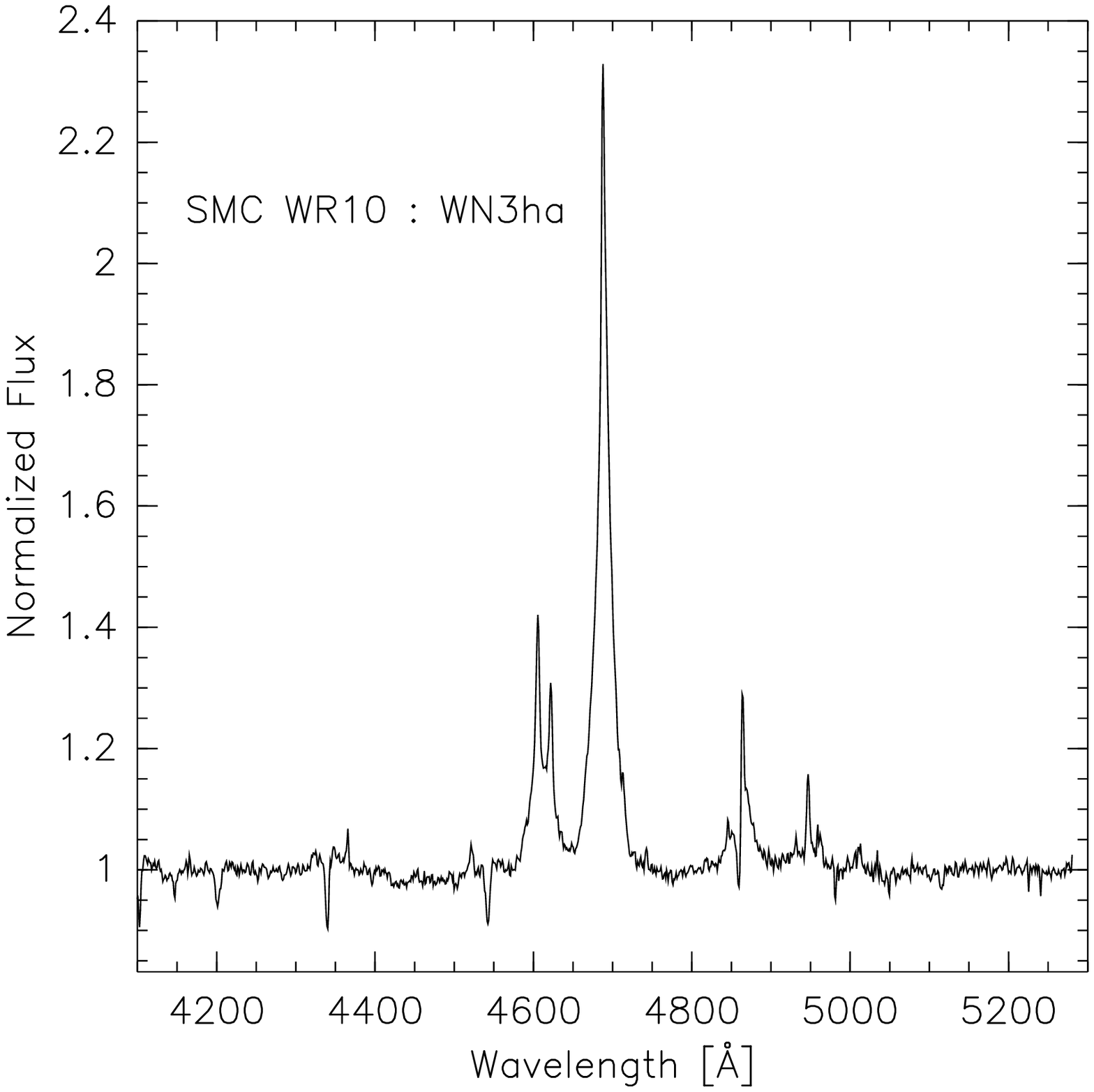}{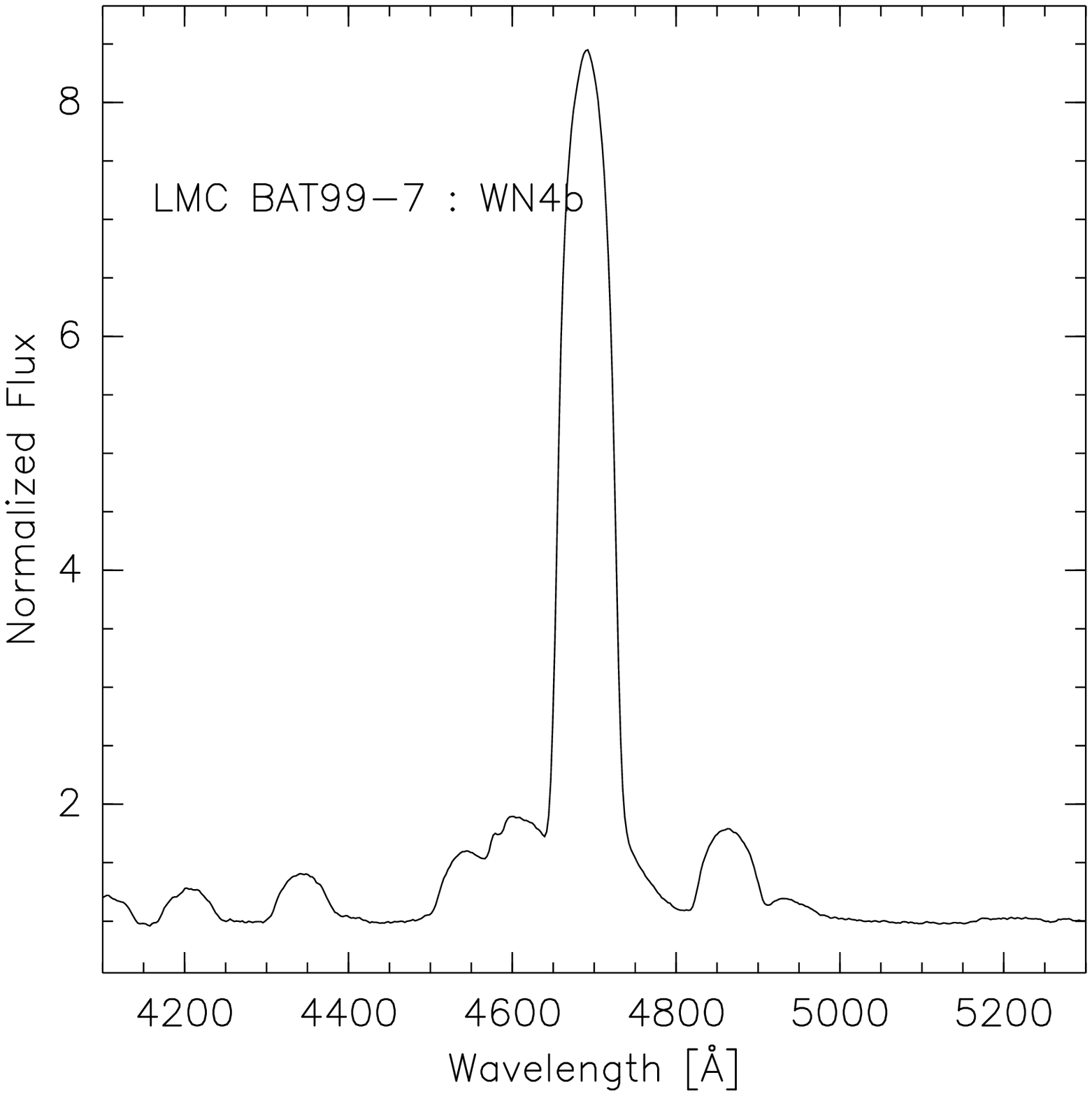}
\caption{Spectrum of SMC-WR10 and LMC BAT99-7 \citep[see][for the WR names in the LMC]{bat99}. The two stars are hot early-type WN stars, but their spectra are completely different. The main origin of this difference is the large amount of hydrogen in SMC-WR10 (or vice versa the lack of such hydrogen in BAT99-7). In the context of H-content, the first star is clearly a late-type WN star, called "eWNL", while the star on the right is a true hydrogen-free eWNE star (see text).}
\label{wne_not_wne}
\end{figure}

Hydrogen has also a direct impact on our understanding of the WR classification (especially of the WN sequence), as shown in Fig.~\ref{wne_not_wne}, where two WNE stars are plotted. The differences are not only qualitative. For instance, \citet{Hamann-etal-1995} noted the physical property jump in their models of WN stars between the stars with broad lines \citep[i.e. using the "b" label, following][]{Smith-etal-1996} and the other (weak lined) WN stars. Moreover, \citet{Smith-Maeder-1998} have shown that the notations "ha-h-(h)-o" in the Smith et al. classification scheme are probably an evolutionary sequence. Finally, \citet{Foellmi-etal-2003b} have shown a clear distinction between the WN stars with broad lines (such as in BAT99-7, see Fig.~\ref{wne_not_wne}) and those labelled "ha/h/(h)" among the 61 WNE stars in the LMC: only two of them have a mixed subtype "b(h)".

\section{Massive-star classification.}

All these evidence led \citet{Foellmi-etal-2003b} to propose a new classification for massive stars that is complementary to that of \citet{Smith-etal-1996}. This evolutionary classification is recognizable by putting an "e" in front of the class, and is based this time on the hydrogen content, i.e. the broad line criterion becomes the principal classification criterion. To summarize, the stars classified as O, Of, WN6 and WN7 are called "eO" since they are certainly core hydrogen-burning \citep[see][for a discussion on the very massive WN6 and WN7 stars]{Foellmi-etal-2003b}. The transition objects RSG, LBV, WN9-11 are called "eOW". The WR stars with hydrogen (whatever their ionization subclass, except WN6 and WN7) are eWNL, while the real hydrogen-free WN stars (i.e. with broad lines) are labeled "eWNE". With this classification, models and observations agree \citep[see e.g.][ and also Fig.~\ref{wne_not_wne}]{Foellmi-etal-2003b,Meynet-Maeder-2005}. While the ionization subclass is thought to be metallicity-dependent \citep[see e.g.][]{Crowther-2000}, the evolutionary classification is valid at all $Z$, following which the H-rich WN stars in the SMC are "hot eWNL stars". 

\begin{figure}[!t]
\plotone{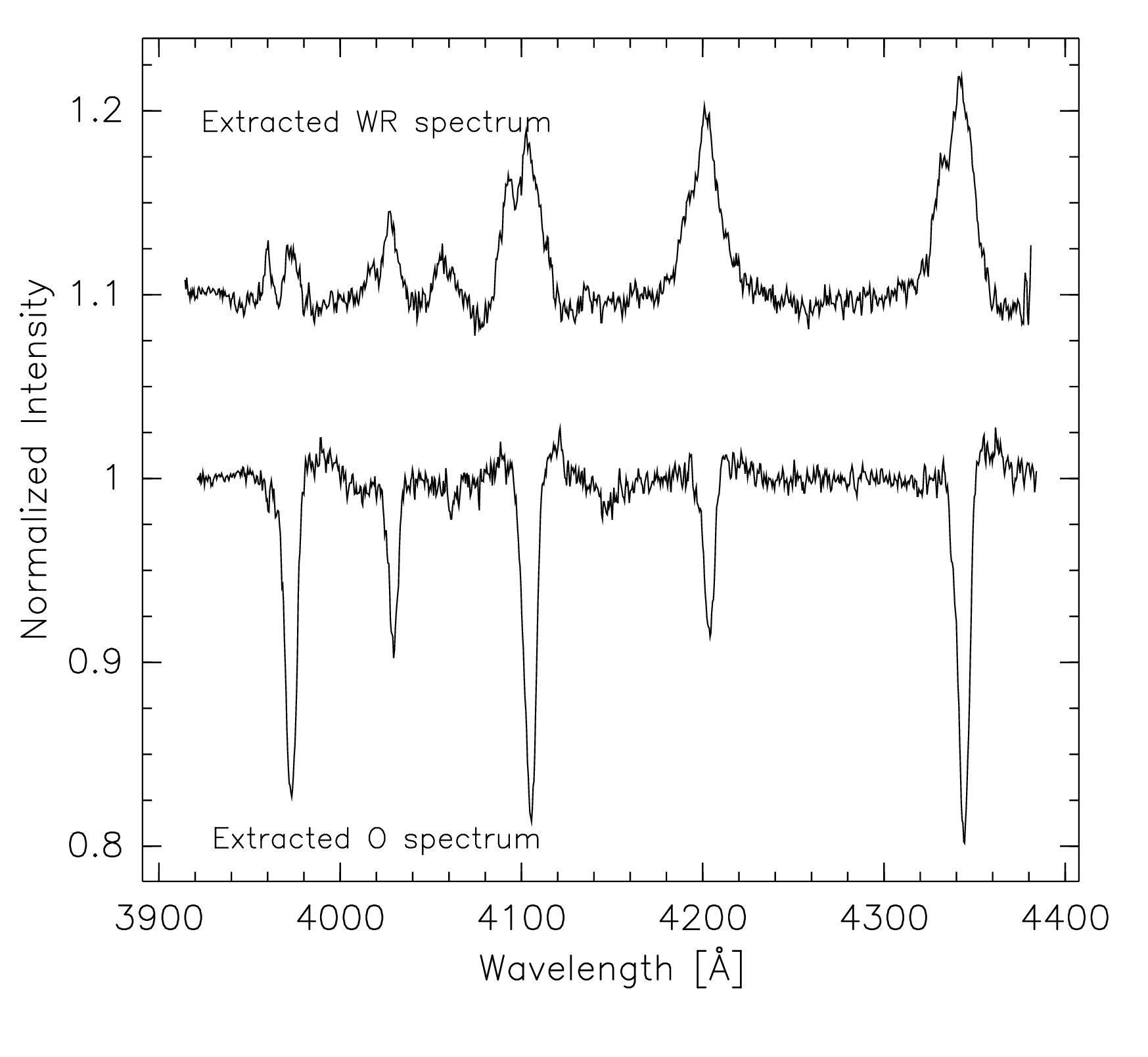}
\caption{Extracted spectra of the O and WR component in the LMC eclipsing binary BAT99-129 \citep{bat99}. It can be easily seen that blueshifted absorption lines (which follow the WR orbital motion) appear on top of the emission lines in the WR spectrum. See \citet{Foellmi-etal-2005-astroph} for details. }
\label{bat129}
\end{figure}

Recently, \citet{Foellmi-etal-2005-astroph} have found a significant amount of hydrogen in a short-period {\it binary} in the LMC (see Fig.~\ref{bat129}). The WR component is classified WN3ha. According to a first modeling (P. Crowther, private communication), the amount of hydrogen is similar to that found in SMC stars, i.e. the H/He ratio (by number) is about unity, while the temperature reaches $T_{eff} \sim 90 kK$! The evolutionary classification is therefore also applicable to binary stars.

Hot eWNL stars were also found in the LMC \citep[about one third of the WNE population:][]{Foellmi-etal-2003b}, and more recently, one such star in our Galaxy, in a region where the ambient metallicity may resemble that of the LMC: WR3, classified as WN3ha \citep{Marchenko-etal-2004}. The variation of the fraction of such stars with metallicity strongly suggests that they are the dominant type of WR stars in low-metallicity environments, i.e. mainly at high redshift. It is therefore relevant to ask: could they be the progenitors of long-soft Gamma-ray bursters (GRB)?

\section{Progenitors of GRBs}

We recall here the main theoretical ingredients to form a GRB, following \citet{MacFadyen-Woosley-1999}: (1) A massive core is needed to form a black hole, (2) enough angular momentum must remain in the core to produce bipolar jets and (3) the star must have lost its hydrogen envelope to allow the radiation to reach the surface and escape. The third point implies that if a WR star at low metallicity is indeed to become a progenitor of a GRB, it must go through an H-free WC phase. 

Following the models of rotating WR stars by \citet{Meynet-Maeder-2005}, at low metallicity, only stars with an initial mass above or equal to 60 $M_{\odot}$ will go through a non-negligible WC phase. Very few stars will be massive enough to reach the WO stage. Thus, it is very unlikely that the hot eWNL stars are good GRB candidates ones, and the GRB progenitors must be found among higher-mass stars. This idea is reinforced by the recent modeling of \citet{Hirschi-etal-2005-astroph}, who claim that basically only the stars that reach the WO stage will be capable of producing GRBs. 

However, the supernova SN1998E associated with GRB980910 showed hydrogen in its spectrum \citep{Rigon-etal-2003}. Moreover, recently \citet{Starling-etal-2005a} also found hydrogen in significant quantities in the afterglow of GRB021004.  As suggested by the analysis of WR3 in our Galaxy \citep{Marchenko-etal-2004}, the absorption lines form relatively close to the stellar core. It is likely that the absorbing material is then gradually accelerated to the terminal wind velocity. Since we have shown that it is not unusual to see a significant amount of hydrogen in single and binary WN stars at low metallicity, it might not be too surprising to observe high-velocity hydrogen absorption features in GRB afterglows, such as described by \citet{Starling-etal-2005a} on GRB 021004.

On the other hand, \citet{vanMarle-etal-2005-astroph} argue that the presence of an intermediate velocity component in the afterglow of GRB 021004 implies that the WR phase was short, i.e. that the WR shell was still intact when the star exploded. This implies that the initial mass of the progenitor must have been small (i.e. about 25 $M_{\odot}$), since the smaller the initial mass, the shorter the WR phase \citep[see Fig. 9 in][]{Meynet-Maeder-2005}. But such stars have a very short H-free WC stage. Although this is possibly in contradiction with the requirement of no hydrogen left in the atmosphere \citep{MacFadyen-Woosley-1999}, it favors the idea of hot eWNL stars as GRB progenitors.

It is interesting to note that theoretical models seem to produce two very different types of GRB progenitors. On one hand, models require H-free progenitors and \citet{Hirschi-etal-2005-astroph} claim that single (initially massive) WO type stars are the best GRB candidates., On the other hand, the presence of hydrogen in GRB afterglows, the results of \citet{vanMarle-etal-2005-astroph} and the fraction of hot eWNL point toward lower mass WR stars. Could the solution come from short-period binaries? Possibly, the only way to reconcile the need for a non-negligible WC phase and low initial mass (i.e. $<$ 60 $M_{\odot}$) could be in an {\it interacting} short-period binary, such as BAT129 in the LMC \citep{Foellmi-etal-2005-astroph}. In that context, the unique WO 16-day binary star of the SMC (WR8) deserves probably much more attention \citep[see e.g.][]{Bartzakos-etal-2001}.

\acknowledgements C.F. thanks P.A. Crowther for having carried out a first modeling of the spectrum of BAT129.

\end{document}